\documentclass[aps,prl,twocolumn,groupedaddress]{revtex4-1}
\usepackage{graphicx}
\usepackage{natbib}
\usepackage{keyval}
\usepackage[justification=RaggedRight]{caption}[2008/04/01]
\usepackage{subfig}

\begin{document}

%\preprint{}

\title{On the Properties of Hydrogen Terminated Diamond as a Photocathode}


\author{J. D. Rameau$^{1}$}
\author{J. Smedley$^{1}$}
\author{E. M. Muller$^{2}$}
\author{T. E. Kidd$^{1, 3}$}
\author{P. D. Johnson$^{1}$}

%\thanks{}
\affiliation{$^{1}$Brookhaven National Laboratory, Upton, NY, 11973, USA}
\affiliation{$^{2}$Stony Brook University, Stony Brook, NY, 11794, USA}
\affiliation{$^{3}$University of Northern Iowa, Cedar Falls, IA, 50613-01500, USA}

\date{\today}

\begin{abstract}
Electron emission from the negative electron affinity (NEA) surface of hydrogen terminated, boron doped diamond in the [100] orientation is investigated using angle resolved photoemission spectroscopy (ARPES). ARPES measurements using 16 eV synchrotron and 6 eV laser light are compared and found to show a catastrophic failure of the sudden approximation. While the high energy photoemission is found to yield little information regarding the NEA, low energy laser ARPES reveals for the first time that the NEA results from a novel Franck-Condon mechanism coupling electrons in the conduction band to the vacuum. The result opens the door to development of a new class of NEA electron emitters based on this effect.
\end{abstract}

%\keywords{}

\maketitle

The electron affinity of a semiconductor surface is defined as $\chi\equiv E_{vac}-E_{CBM}$ where $E_{vac}$ is the ``binding energy" of the vacuum level and $E_{CBM}$ is that of the conduction band minimum. Semiconductors for which the electron affinity is \textit{negative}, meaning that $E_{vac}$ lies within the band gap, are prized for their unique potential as advanced electron emitters\cite{diamondoid}. This is because a negative electron affinity (NEA) surface generally leads to very efficient emission of electrons into the vacuum. The emitted beams also often possess characteristics such as narrow energy spread and low emittance that are highly desirable for use in devices such as electron microscopes, photoinjectors for free electron lasers and the recently demonstrated diamond electron amplifier\cite{amplifier}. Most NEA surfaces are difficult to prepare and maintain, usually requiring the deposition of alkali metals on a specially prepared surface and which exhibit a subsequently short operational lifetime. The [100] surface of hydrogen terminated diamond (H:C[100]) has proven to be a remarkable exception. A $\chi$ on the order of -1 eV is repeatedly obtainable by a simple annealing procedure and the surface is robust against exposure to air and other contaminants. H:C[100] achieves this feat through the combination of its large band gap $E_{g}=5.5$ eV and robust H terminated surface, stable up to 800 $^{o}$C, that drives $E_{vac}$ into the band gap.

Interest in diamond has been recently rekindled due to the increasing availability of economically viable, device-quality, synthetic crystals as well as the obvious advantages of functionalizing a material possessing such exceptional thermal, electrical and mechanical properties. It is therefore remarkable that no agreed upon mechanism for NEA emission from H:C[100] has been established. In the following Letter we present angle resolved photoemission spectroscopy (ARPES) measurements showing how H:C[100] attains a useful NEA surface by means of strong electron-optical phonon (e-ph) coupling and a breakdown of the sudden approximation. This mechanism is potentially germane to not only NEA emission but also to a range of recently demonstrated diamond based devices\cite{amplifier, dlaser, led}.

The vast majority of PES experiments performed on H:C[100] have been carried out using photon energies well in excess of $E_{g}$(\cite{example,pate} for example). A notable exception to this trend was the recent laser PES experiment\cite{shinprl} carried out at a photon energy of 7 eV. This experiment yielded information on the strength of the electron-optical phonon (e-ph) coupling with an eye towards explaining the mechanism by which boron doped diamond (BDD) traverses the insulator-metal-superconductor phase diagram. However, no information about NEA emission was acquired. A PES study on H:C[111] \cite{pate,pateups} using an incoherent light source with photon energies between 5 and 6 eV suggested a mechanism of phonon assisted exciton desorption, though the relation between emission from the [111] and [100] surfaces is unclear.

The present ARPES experiments were performed at beamline U13UB of the National Synchrotron Light Source. Synchrotron (SR) based photoemission was carried out at 16 eV photon energy. Laser light of 6.01 eV (208 nm) was generated as the fourth harmonic of a mode-locked Ti:Sapphire laser\cite{me} and supplied simultaneously to the same endstation. The synchrotron and laser beams were \textit{p} and \textit{s} polarized, respectively, relative to the sample surface. The endstation was equipped with a Scienta SES-2002 hemispherical electron spectrometer and maintained at a base pressure of $5\times10^{-11}$ Torr. The overall energy resolution of the experiment was set to 10 meV for both the SR and laser experiments to facilitate rapid data collection. The angular resolution was 0.1 degrees. The Fermi energy $E_{F}$ was referenced at both photon energies to a polychrystaline gold film evaporated onto the sample holder and in simultaneous electrical contact with both the diamond sample and ground. Acceleration of electrons in vacuum for the SR PES experiment was performed using a precision Valhalla 2701C voltage reference.

The sample used in the experiment was a device quality, single crystal type IIb diamond obtained from Element Six\cite{e6}. The sample, grown by chemical vapor deposition (CVD), measured 3.5 mm$^{2}$ by 500 $\mu$m thick and was boron doped to the level of $n_{B}=2\times10^{19}$ cm$^{-3}$ (.011\% B/C). No sample charging was observed during the experiment. A monolayer of monatomic hydrogen was deposited \textit{ex situ} upon the polished [100] surface of the sample prior to the experiment\cite{amplifier}. The sample was annealed \textit{in situ} to 400 $^{o}$C for about one hour in a vacuum of $10^{-7}$ Torr to remove any adsorbed water and then allowed to return to room temperature before commencing the ARPES experiment. Low energy electron diffraction performed at the end of the experiment revealed an unreconstructed surface.

\begin{figure}
  \includegraphics[width = 75mm, bb = 118 245 481 546]{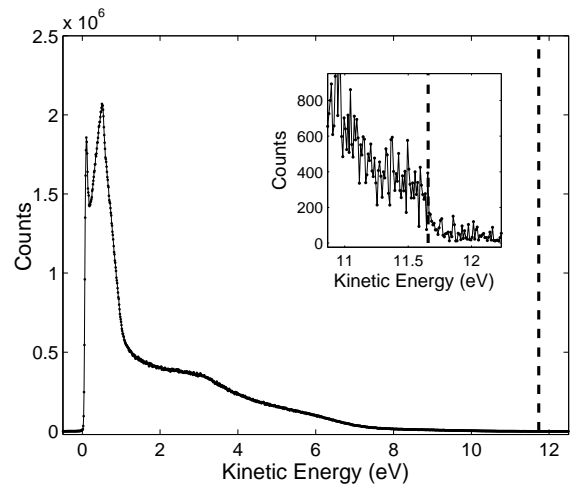}\\
  \captionsetup[figure]{justification=justified}
  \caption{Density of states for H:C(B)[100] at 16 eV photon energy. $E_{F}$ is marked by the dashed line at 11.73 eV yielding a work function $\phi=4.27$ eV. Inset: a zoom of the near-$E_{F}$ spectrum. The kinetic energy has been corrected for the external bias and the offset imparted by the analyzer. Lines connecting points are a guide to the eye.}\label{synchro}
\end{figure}

To put the laser ARPES in the proper context we performed an SR experiment similar to those previously reported. In order to access kinetic energies down to near zero the sample was biased relative to ground by -10 V. A typical SR integrated energy distribution curve (EDC) at normal emission is shown in Figure \ref{synchro}. The spectrum resulting from SR photoexcitation is qualitatively similar to those reported previously\cite{pateups,upsexample,example}. While the resistivity measurement shows the sample to be nominally insulating, a weak Fermi cutoff (inset of Fig. \ref{synchro}) was detected in the density of states. As in previous photoemission experiments we observe a bright, narrow peak at nearly zero KE separated from another, broader peak by $\sim0.5$ eV. These have previously been shown to be common features of the NEA surface of H:C[100]. The bulk of the SR spectrum consists of valence band photoemission usually understood in the context of the sudden approximation, in which electrons are excited directly from an occupied state to a free electron final state in the vacuum\cite{hufner} upon absorbing a photon. The remainder of the photoemission spectrum consists of various secondary electrons appearing as an energy dependent background, not all of which can be associated with NEA emission. Even in the best case the emission characteristic of the PES spectrum shown in Fig. \ref{synchro} consists of a superposition of both direct and secondary emission, the varying components of which are difficult to disentangle.

\begin{figure}
  \includegraphics[width = 75mm, bb = 114 239 488 532]{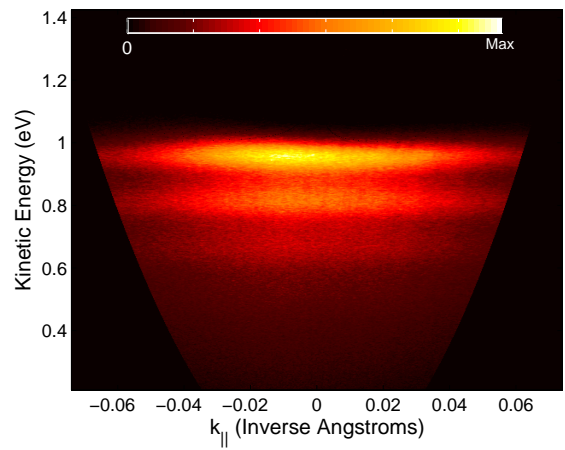}\\
  \caption{ARPES spectrum acquired using the 6.01 eV laser. The color bar shows relative intensity. The overall ``bowl" shape of the spectrum results from scaling between emission angle and $k_{||}$. (Color online)}\label{laserspec}
\end{figure}

In Fig. \ref{laserspec} we present the laser ARPES spectrum of H:C(B)[100] taken in the absence of a bias on the sample. The laser ARPES spectrum is characterized by a set of well defined intensity peaks spaced at regular intervals in energy. We also observe a large gap in the spectrum between the highest energy peak at $\sim1$ eV KE and $E_{F}$, located at 1.662 eV as determined by the gold reference. The lack of reference to $E_{F}$, in addition to the lack of a Fermi edge or any Fermi edge replicas\cite{shinprl}, signifies that the laser generated photoelectrons emanate from the conduction band due to the NEA surface.

\begin{figure}
  \includegraphics[width = 75mm, bb = 114 242 488 548]{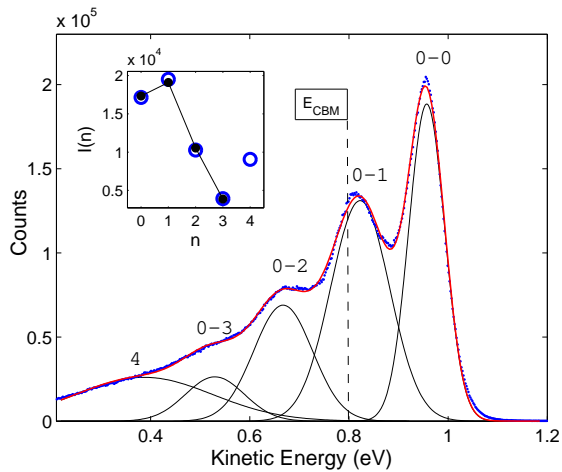}\\
  \caption{EDC integrated between $\pm0.02$ \AA$^{-1}$. Points (blue) show data. The line through the points (red) is the total fit. The decomposition of the fit into constituent Gaussian curves (black) is shown for comparison. Inset: integrated intensities of Gaussian components verses peak number $n$ (open circles) along with FC fit (closed circles) for $n$ = 0 to 3. Lines between points are a guide to the eye. The dashed line indicates the position of $E_{CBM}$ given by the FC fit. (Color online)}\label{fits}
\end{figure}

In Fig. \ref{fits} we plot the EDC integrated over a narrow momentum range about $k_{||}=0$ along with fits to the spectrum. The spectrum is well fit down to 0.21 eV by a series of Gaussian peaks with \textit{no background} over most of the range. The mean peak spacing is found to be $142 \pm 5$ meV. We assign the secondary peaks as vibrational sidebands of the main peak at $\sim1$ eV. The appearance of such pronounced, discreet quantum electronic states at room temperature is highly unusual in a three dimensional solid. These sidebands arise in the presence of a strong e-ph coupling. The apparent broadness in $k$ arises from the large number of final momentum states imparted to escaping electrons during the emission process. Comparing Figs \ref{synchro} and \ref{fits} we thus find that the laser and SR ARPES results are profoundly different, heralding a breakdown in the sudden approximation.

Following Ref. \cite{shinprl} we fit the integrated intensities of peaks 0 through 3 with the Poisson distribution characteristic of phonon emission due to the Franck-Condon (FC) principle such that $I(n)\propto e^{-g} g^{n}/n!$ where $g$ is the effective e-ph coupling constant\cite{mahan}, $n$ is the peak index (starting at $n=0$) and $I(n)$ is the integrated intensity of each peak. The analysis is clearly facilitated by the lack of complications arising from the presence of a Fermi edge or replicas thereof. This formulation of the intensity distribution is valid at room temperature because the observed phonon energy is far too high to be thermally excited. Optical phonon \textit{emission} is thus the dominant process. The  integrated intensities thus extracted, along with the FC fit, are plotted in the inset of Fig. \ref{fits}. Surprisingly we find that at room temperature $g=1.1\pm0.01$, a value in agreement with that obtained previously on a superconducting BDD of more than two orders of magnitude higher doping. Peak 4, which does not follow this trend, is possibly associated with the Fano effect\cite{fano, fanoraman}.

What is the significance of observing this effect in the NEA emission of diamond? It must be recalled that diamond has an indirect band gap, with the conduction band minimum (CBM) lying approximately three quarters of the way to the zone edge, at the $\Delta$ point. Photoexcitation to the CBM entails a second order process involving the absorption of a photon followed by emission of a phonon of energy $\hbar\Omega_{0}$ and finite momentum $k_{CBM}$. Under low energy laser excitation this process occurs in the bulk of the sample. Because the CBM lies at a large absolute value of $k_{||}$ in the first Brillouin zone\cite{bz}, very low energy emission of electrons from the sample surface, from the CBM, is kinematically inhibited\cite{hufner}\cite{pate}. Direct emission of valence band electrons at low $k_{||}$ is suppressed by the absence of a well defined free electron final state. Nevertheless strong photoemission is observed, at normal emission on the [100] surface, in the absence of any zone center conduction band states.

\begin{figure}
  \includegraphics[scale = .4, bb = 8 234 590 483]{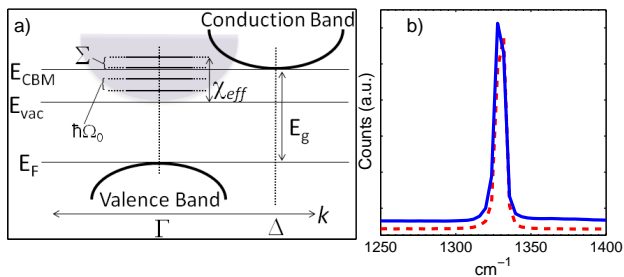}\\
  \caption{a) Sketch of the effective electronic structure of H:C[100]. Shaded region denotes kinematically accessible final states. b) Micro Raman spectra of the Boron doped sample, (blue) solid line, offset vertically by 400 for clarity, and an undoped CVD diamond, (red) dashed line. The Raman laser had a wavelength of 532 nm. (Color online)}\label{raman}
\end{figure}

The FC principle allows a natural physical explanation for the NEA emission ubiquitously observed normal to the H:C[100] surface. Electrons excited in the bulk transition to the conduction band rapidly thermalize to the vicinity of $(k_{CBM}, E_{CBM})$. As mentioned above, their high crystal momentum parallel to the surface inhibits direct emission into free electron final states in the vacuum. Additionally, the fact that $E_{CBM}$ lies far above $E_{vac}$ strongly inhibits decay back to the valence band for hot carriers diffused to the surface. Inclusion of phonon dressing in the escape mechanism allows momentum conserving transitions for electrons from $(E_{CBM},k_{CBM})$ to $(E_{k},k_{||}=0)$. Further evidence that this is the case arises from the fact that the optical phonon energy observed in ARPES agrees well with the transverse optical mode observed at $k_{CBM}$ by neutron scattering\cite{warren}. The situation is sketched in Fig. \ref{raman}a. Indeed, Raman measurements on the present BDD, Fig. \ref{raman}b, show a negligible softening of the main emission line around 1325 cm$^{-1}$ (165 meV) from that of a similar undoped CVD diamond, thereby ruling out zone center phonons as the excitation relevant to the observed PES spectrum. We therefore find that the unique, high energy phonon spectrum of diamond plays the key role in the NEA electron emission of H:C[100], for without the FC effect generated by the large e-ph coupling no electron emission into the vacuum would be possible from the CBM for any realistically attainable work function. The emitted electrons and phonons constitute a single many body final state; the observed bands at normal emission are not, properly speaking, virtual states, but composite states of CBM electrons dressed by the e-ph interaction.

The effect observed here is analogous to that causing the appearance of vibrational sidebands in absorption, fluorescence and core level spectroscopies of diatomic molecules, though in this case manifest as electron rather than photon emission. However this represents to our knowledge the first observation of a classic FC principle in action for electronic transitions from excited states of a solid to the vacuum rather than from the true ground state. The peak energies at normal emission are $E_{k}=E_{CBM}+\Sigma-n\hbar\Omega_{0}$ where $E_{vac}\equiv0$ KE, $\Omega_{0}$ is the optical phonon frequency and $\Sigma=g\hbar\Omega_{0}$ is the real part of the electron self energy\cite{mahan}. The self energy has the effect of shifting the maximum KE of emission above $E_{CBM}$ by $g\hbar\Omega_{0}$. Defining the effective NEA as the highest KE \textit{peak} position we find that $\chi_{eff}=E_{vac}-(E_{CBM}+\Sigma)=-0.955\pm0.005$ eV.

At room temperature, a reduction of boron doping by more than two orders of magnitude relative to the previous work has no observable effect on the effective e-ph coupling, $g$. This finding indicates the increase of $T_{c}$ observed in this system, with increasing doping, is mostly due to the rising electronic density of states at the Fermi level. Similarly, the e-ph coupling in undoped diamond, and associated NEA properties of its H terminated surface, likely derive from the same e-ph coupling. The observation of a Fermi edge in the SR PES spectrum appears to signify passage across the metal-insulator transition (MIT) at a doping level somewhat lower than previously reported, though in good agreement with the Mott criterion\cite{pps}. This observation is possibly due to upwards band bending at the surface, the region SR PES is most sensitive, versus the more bulk sensitive laser PES. Regardless, observation of metallic behavior in the SR PES and essentially semiconducting behavior in the laser PES, with no Fermi edge, exemplifies violation of the sudden approximation to an extent not previously reported. Lastly we note that the coherent excitation of a non-thermal population of a single optical phonon mode carrying large momentum indicates that diamond may be the ideal material for constructing a robust, monolithic SASER\cite{saser1, saser2, saser3} operable at room temperature. Realization of such a device would constitute a major advance in the nascent field of applied phononics.

The laser induced electron spectrum of H:C(B)[100] is highly unconventional in that it originates entirely from the conduction band of the material in a ``single photon" photoemission experiment. The important role strong e-ph coupling plays in determining this spectrum is consistent, for instance, with recent theoretical investigations of the phonon induced renormalization of the bare band structure of diamond\cite{cohen}. We have demonstrated that laser ARPES is the ideal tool for examining the properties of NEA surfaces as the resulting spectrum lacks the complications that arise in typical high energy PES experiments. NEA emission from BDD near the MIT is shown to result from strong e-ph coupling and a concomitant breakdown of the sudden approximation for photoemission manifest as a set of many body final states in the band gap not directly related to the bulk band structure. Key to demonstrating this was the ability to use both laser and synchrotron radiation on one sample at a single facility. Finally, not only does the revelation at long last of how H:C[100] attains NEA electron emission explain how devices such as the diamond amplifier and field emitters based on this effect actually work but firmly establishes diamond as a class of NEA material distinct from ``direct gap" emitters such as GaAs.

\begin{acknowledgments}
We thank Philip Allen for illuminating discussions. This research was supported by the U.S. Department of Energy, Basic Energy Sciences, Materials Sciences and Engineering Division and performed at the Center for Functional Nanomaterials and the National Synchrotron Light Source, DOE-BES user facilities at Brookhaven National Laboratory, and by DOE grant DE-FG02-08ER41547. T.E. Kidd was supported by the Iowa Office of Energy Independence Grant No. 09-IPF-11.
\end{acknowledgments}

%Merlin.mbs v4.21 2009-07-09.
%

\end{document}